# The role of refractive index in metalens performance

ELYAS BAYATI [1], ALAN ZHAN [2], SHANE COLBURN [1], ARKA MAJUMDAR[1,2, *]

[1]Department of Electrical Engineering, University of Washington Seattle, WA-98195
[2]Department of Physics, University of Washington, Seattle, WA-98195
*Corresponding author: arka@uw.edu

**Sub-wavelength diffractive optics, commonly known as metasurfaces, have recently garnered significant attention for their ability to create ultra-thin flat lenses with high numerical aperture. Several materials with different refractive indices have been used to create metasurface lenses (metalenses). In this paper, we analyze the role of material refractive indices on the performance of these metalenses. We employ both forward and inverse design methodologies to perform our analysis. We found that, while high refractive index materials allow for extreme reduction of the focal length, for moderate focal lengths and numerical aperture (<0.6), there is no appreciable difference in focal spot-size and focusing efficiency for metalenses made of different materials with refractive indices ranging between n= 1.25 to n=3.5.**

## 1. Introduction

Dielectric metasurfaces, two-dimensional quasi-periodic arrays of subwavelength scatterers have recently emerged as a promising technology to create ultra-thin, flat and miniature optical elements [1]. With these sub-wavelength scatterers, metasurfaces shape optical wavefronts, modifying the phase, amplitude, and/or polarization of incident light in transmission or reflection. Many different optical components such as lenses [2, 3], focusing mirrors [4], vortex beam generators [5, 6], holographic masks [7, 8], polarization optics [9, 10] and freeform surfaces [11] have been demonstrated using metasurfaces. While the sub-wavelength structuring provides the necessary phase-shift for light manipulation, the material degrees of freedom also play an important role in determining the performance.

Metasurfaces initially relied on deep-subwavelength metallic structures and operated at mid-infrared frequencies [1]. The large absorption loss in metals made it difficult to create high-efficiency metasurface devices in the visible and near-infrared (NIR) wavelengths. This motivated the fabrication of metasurfaces using dielectric materials because of the low optical loss of dielectrics at visible and NIR wavelengths. While initial research focused on higher-index amorphous silicon (Si) [3, 12] at NIR wavelengths, recently materials with lower refractive index, such as titanium oxide (TiO$_2$) [13], gallium nitride (GaN) [14], and silicon nitride (SiN) [15] have been used to create metasurfaces operating at visible wavelengths. Based on the empirical Moss relation $n^4 \sim 1/E_g$ [16] with refractive index $n$ and the electronic bandgap $E_g$, we expect that a large optical transparency window necessitates the material refractive index to be lower. Hence, to create metasurfaces at shorter wavelengths, we have to rely on materials with lower refractive index. In decreasing the refractive index, however, it is unclear what effect there will be on device performance. Recently, the efficiency of a periodic meta-grating was analyzed at optical frequency as a function of the material refractive index [17]. They reported that for large deflection angles the efficiency decreases with lower refractive index, but for low deflection angles there is no significant difference in efficiency of transmissive thick meta-gratings made of different materials. While this analysis with periodic structures can help qualitatively understand the performance of a metasurface lens (metalens) with quasi-periodic arrangements of scatterers, a systematic and quantitative evaluation of material selection for metalenses is currently lacking. It is unclear what the minimum required dielectric contrast is to achieve high-efficiency and high numerical aperture metalenses. Answering this is vital for understanding the capabilities, limitations, efficiency and manufacturability of metalenses over a specific wavelength range. We note that, the effect of refractive index is explicit in the lens maker's formula [18] for a refractive lens:

$$\frac{1}{f} = (n-1)(\frac{1}{R} - \frac{1}{R'})$$

where $f$ is the focal length, $n$ is the refractive index of the lens, and $R$ and $R'$ are the radii of curvature of the two spherical surfaces of the lens. The angle of refraction, and therefore the focal length depends on the curvature of the lens surface and the material used to construct the lens. However, for metalenses, there is no study or theoretical formula relating the refractive index to the focal length or numerical aperture and the efficiency.

In this paper, we design and analyze metalenses made of materials with wide range of refractive indices to estimate the relationship between refractive index and performance of metalenses. We analyze metalenses operating in the near infrared spectral regime (λ = 1550nm) in terms of efficiency and full width at half maximum (FWHM). We consider six different dielectric materials: Si (n = 3.43) [3], TiO$_2$ (n = 2.4) [12], GaN (n = 2.3) [13], SiN (n = 2.0) [14], SiO$_2$ (n=1.5) [19] and an artificial material with a refractive index of 1.2. The index range under n=2 is of particular importance as large-scale printable photonics technology requires the refractive index to be near 1.5 [20] and promising for low-cost manufacturing of metasurfaces. First, we used a forward design technique based on rigorous coupled-wave analysis (RCWA) [21, 22] followed by finite-difference time-domain (FDTD) simulations [3, 14]. We compare the focusing efficiency and FWHM at the focal-plane as a function of the numerical aperture for different materials. We then employed inverse electromagnetic design based on generalized Mie scattering theory and adjoint optimization [23] to calculate the dependence of metalens performance and FWHM at the focal-plane on refractive indices between 1.25 and 3.5.

## 2. Forward Design Method

The main building block of a metalens is a scatterer arranged in a subwavelength periodic lattice (with a period $p$). Here, we assume the scatterers to be cylindrical pillars, arranged in a square lattice, as shown in Figure 1. Since we have sub-wavelength periodicity in a metalens, only the zeroth-order plane wave propagates a significant distance from the metasurface and other higher order diffracted plane waves are evanescent [24]. This makes metalenses more efficient compared to other diffractive optics.

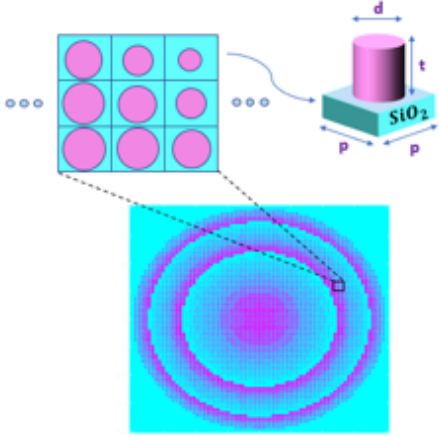

Figure 1. Schematic of a metalens and its lattice structures: a lattice with periodicity $p$ can be formed by using cylindrical pillars (with diameter $d$ and thickness $t$) on top of a silicon dioxide substrate, arranged in a square lattice. By varying the radius of the cylindrical pillars, we can impart different phase-shifts.

Forward design of a metalens involves selecting the appropriate spatial phase profile for the specific optical component, arranging the scatterers on a subwavelength lattice, and spatially varying their dimensions. To have an arbitrary transmission phase profile, phase shifts of scatterer should span the 0 to $2\pi$ range while maintaining large transmission amplitudes. In our simulation, we used the phase-profile of metalens as:

$$\phi(x,y) = \frac{2\pi}{\lambda}\left(\sqrt{x^2 + y^2 + f^2} - f\right)$$

We discretize this continuous spatial phase profile onto a square lattice with periodicity $p$, giving us a discrete spatial phase map with different phase values. We then quantize the phase profile with ten linear steps between 0 and $2\pi$, corresponding to ten different pillar radii. For each value of this new discrete spatial phase profile, we find the radius of the pillar that most closely reproduces that phase and place it on the lattice.

The complex transmission coefficient of zeroth-order plane wave depends upon the lattice periodicity $p$, scatterer dimensions (both the diameter $d$ and thickness $t$), and refractive index $n$. Using RCWA, we calculate the phase and transmission amplitude of the scatterers as a function of $d/p$ for different materials assuming periodic boundary condition (Figure 2). For different refractive indices, we can find several sets of thickness $t$ and lattice periodicities $p$, which provide the whole 0 to $2\pi$ phase shift range under varying diameters while maintaining high transmission amplitude (transmissivity ~1). Some resonant dips in transmission are observed, which can be attributed to the guided mode resonances [25]. Metasurface parameters including lattice periodicity $p$ and thickness $t$ for each material are shown in Table 1. As we are comparing different materials, we chose these parameters to maintain the same aspect ratio across simulations, in this case selecting $t/p \sim 1.6$. For our artificial material with refractive index of 1.2, however, in order to cover the whole 0 to $2\pi$ phase shift range, we need to increase the thickness. To keep the same aspect ratio, we cannot maintain sub-wavelength periodicity. Hence, for n=1.2, we assume an aspect ratio of 3.6 to get the maximum possible phase shift. We assume the substrates for all materials to be SiO$_2$ with a thickness $t_{sub} = \lambda$.

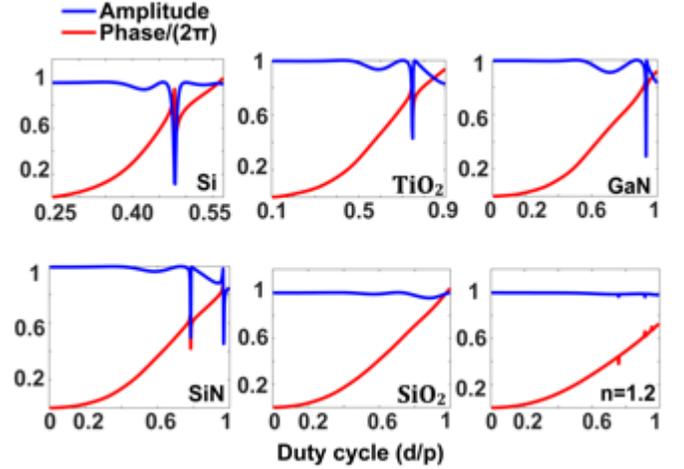

Figure 2. Amplitude and phase of the transmitted light through a scatterer: using RCWA, we calculate the transmission properties (red: phase delay; blue: transmission amplitude) as a function of the ratio of the pillar diameter and periodicity. We kept the aspect ratio, i.e., the ratio between the thickness and the periodicity same for all materials except for n=1.2 to compare their efficiency.

Using the parameters obtained from RCWA, we designed arrays of nanopillars and simulated the metalenses using Lumerical FDTD solutions. The pillar diameters corresponding to resonances in Figure 2 are excluded when designing the metasurfaces to get higher efficiency. We analyzed the performance of the metalenses in terms of FWHM and focusing efficiency for different focal lengths ($5 - 200 \mu m$). The diameter of the metalenses is kept constant at 80 $\mu m$.

Table 1. Metasurface parameters including lattice periodicity $p$ and thickness $t$ for each refractive index, which are used in forward design method

| Refractive index ($n$) | Si ($n = 3.43$) | TiO$_2$ ($n = 2.4$) | GaN ($n = 2.3$) | SiN ($n = 2$) | SiO$_2$ ($n = 1.5$) | ($n = 1.2$) |
|---|---|---|---|---|---|---|
| Periodicity ($p$)($nm$) | 775 | 790.5 | 759.5 | 930 | 1372 | 1395 |
| Thickness ($t$)($nm$) | 1240 | 1317 | 1240 | 1550 | 2290 | 5022 |

The FWHM of the focal spot is shown in Figure 3a as a function of numerical aperture, where the solid black curve is the FWHM of a diffraction limited spot of a lens with the given geometric parameters. There is no appreciable difference in the FWHM across the range of simulated indices, except for n=1.2, where the FWHM does not decrease at very short focal lengths. We define the focusing efficiency as the power within a radius of three times the FWHM at the focal plane to the total power incident upon the lens [3, 15]. Figure 3b shows the focusing efficiency as a function of refractive index of all materials for different numerical apertures. We find that the focusing efficiency decreases with higher numerical apertures, as observed before [3, 15]. At low numerical apertures ($NA < 0.6$), the efficiency of the metalenses is almost independent of the material refractive index. The decrease in the efficiency with increasing numerical aperture, however, is more drastic with lower refractive index, and the efficiency drops more quickly in materials with refractive indices below 1.5.

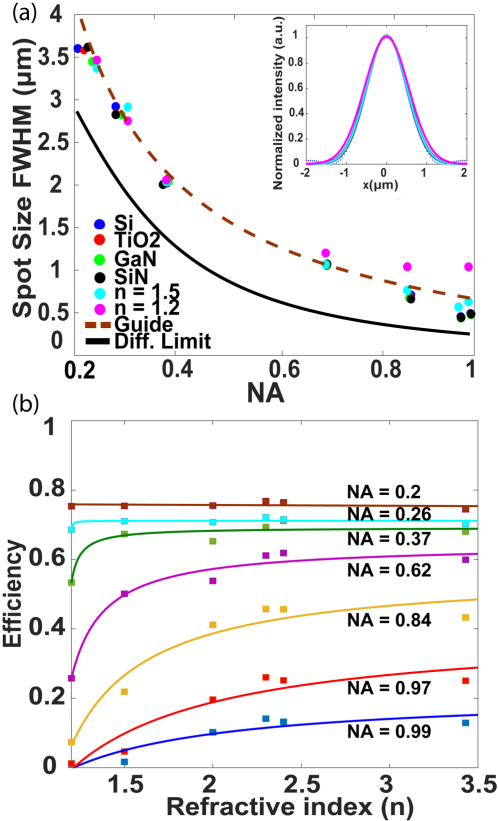

Figure 3. Performance of the metalens designed using forward design methodology: (a) The FWHM of the focal spot is plotted as a function of numerical aperture of the lens. Inset plot of (a) shows cross-section of the beam size for focal length of 50 μm with their Gaussian fit functions. The solid and dashed curves are the diffraction-limited FWHM, and guide to the eye, respectively. (b) The efficiency of metalenses for all materials as a function of refractive index for each numerical aperture using forward design method.

## 3. Inverse Design Method

In the forward design method, we kept the aspect ratio ($t/p$) of the scatterers fixed for different materials. This constraint restricts the design space and makes it difficult to objectively compare metalenses made of different materials. Furthermore, the local phase approximation that we make in going from RCWA to FDTD neglects the coupling between the scatterers. Such coupling is not negligible when the material refractive index is small. We also had to manually inspect the data from RCWA to determine the quality of the parameters, and avoid the resonant dips, before constructing the FDTD simulation. We can circumvent these problems by employing an inverse electromagnetic design methodology, recently developed by our group [23].

Recently, inverse electromagnetic design has been applied to phase profile design [7], single scatterer design [26], beam steering [27], and achromatic metasurface optics [28, 29]. We utilize an inverse design method based on adjoint optimization-based gradient descent and multi-sphere Mie theory which describes the scattering properties of a cluster of interacting spheres. We determine the interactive scattering coefficients for each sphere individually, similar to what Mie theory does for a single sphere [30, 31]. In our inverse design, we do not make any assumption about the size of our scatterers, but we set their periodicity. Consequently, we expect to explore a larger design-space to find the well-suited parameters for our metalens. Our design tool also aims to design the whole metasurface and not just the unit cell, and thus the coupling between scatterers is already included in the design process.

Our inverse design method based on Mie scattering, however, currently only works for spherical scatterers. Hence, in our design, we optimize the radii of different spheres. The radii and periodicity of the metalenses are chosen to avoid any overlap between adjacent spheres. We run the optimization routine up to a fixed number of iterations (in this case 100) to obtain the final metalens. The iteration time of the inverse design method depends on both the particle number and the expansion order of the orbital index $l$. The expansion order here provides the number of spherical basis functions for each particle to include in our simulation [23]. Larger numbers of particles and expansion orders increase the iteration time. As we are interested in sub-wavelength structures to design metalenses, it is important to find a reasonable cut-off for the expansion order to balance the speed of the iteration, and the accuracy of the result. The valid cutoff expansion order ($l_{max}$), which is ultimately determined by the physical size and refractive index of the individual spheres relative to the incident wavelength, is chosen to be 3 in our simulations. Since these scattering properties are determined by the geometric and material properties of the sphere in addition to the wavelength of the incident light, there is a relation between cutoff expansion order and possible periodicity range of the scatterers. Here, the periodicities of the metalenses are chosen in a way such that the contribution from expansion orders greater than 3 are negligible.

Table 2. The periodicity values for ten different refractive indices, which are used in inverse design method

| Refractive index ($n$) | Periodicity ($p$)($nm$) | Refractive index ($n$) | Periodicity ($p$)($nm$) |
|---|---|---|---|
| $1.25$ | 1360 | $2.5$ | 1020 |
| $1.5$ | 1330 | $2.75$ | 976 |
| $1.75$ | 1222 | $3$ | 912 |
| $2$ | 1140 | $3.25$ | 838 |
| $2.25$ | 1122 | $3.5$ | 800 |

We chose ten equally spaced refractive indices between n= 1.25 to n=3.5 for the inverse design. We assume a square periodic lattice, where the spheres with different radii are placed. For all simulations, we assume the spheres are suspended in vacuum, and do not include a substrate. Initially, all the spheres have identical radii. We then allow the sphere radii to vary continuously between 150 nm and half of the periodicity to optimize the figure of merit, which is the intensity at the designed focal point. The periodicities for all refractive indices for the inverse design method are shown in Table 2. The final radius distribution of the optimization process for one metalens using inverse design method is shown in Figure 4. We can see that the result is mostly circularly symmetric as expected for a lens. We attribute the slight asymmetry near the origin to our program running for a fixed number of iterations instead of convergence at a local optimum. Our gradient-based technique solely guarantees convergence to a local optimum, and we manually terminate the algorithm after we reach the desired performance as represented by the FOM. The radius of the designed metalenses is 20 $\mu m$, and five focal lengths between 15 $\mu m$ and 100 $\mu m$ are tested to give us the same number of numerical apertures for better comparison with the forward design.

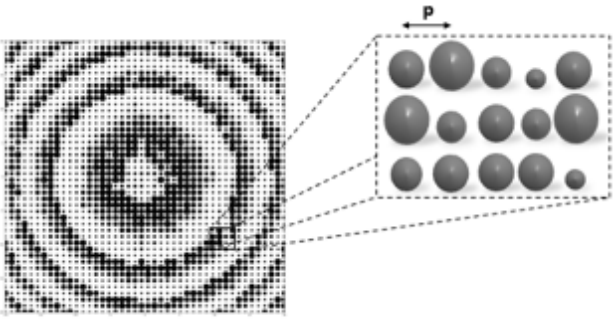

Figure 4. Final radius distribution of a metalens using inverse design method. Spheres are arranged in a square lattice with periodicity $p$. Radii of spheres are allowed to range from 150 nm to the half of this periodicity.

The FWHM of the focal spots of the metalenses are shown in Figure 5a as a function of the numerical aperture, where the solid black curve is the FWHM of a diffraction-limited spot of a lens with the given geometric parameters. Like the forward design method, there is no appreciable difference in the FWHM across the range of simulated indices, except at n=1.25, where the FWHM does not decrease at very short focal lengths. At lower numerical apertures, however, we observe FWHM's smaller than those a diffracted-limited spot. We emphasize that this is not truly breaking the diffraction limit, but rather we attribute this to a larger proportion of the light intensity being located within side lobes as opposed to within the central peak. By shifting power from the central peak to side lobes, beam spot sizes that are less than the diffraction-limited spot size are possible [32]. This shifting of light to the side lobes may have arisen from the defined figure of merit, in which we did not enforce a condition on the beam spot size and side lobe intensity ratio. The efficiencies of these metalenses are plotted as a function of their refractive indices in Figure 5b, where the different numerical apertures are specified with different colors. We find that the focusing efficiency decreases with increasing numerical aperture. According to Figure 5b, there is no significant difference in the focusing efficiency of high and low refractive index materials for low numerical apertures in the range of $NA = 0.37 - 0.2$ (50 – 100 $\mu m$ focal length). For longer numerical apertures, however, such as 0.62 and 0.8, higher refractive index materials provide higher focusing efficiency.

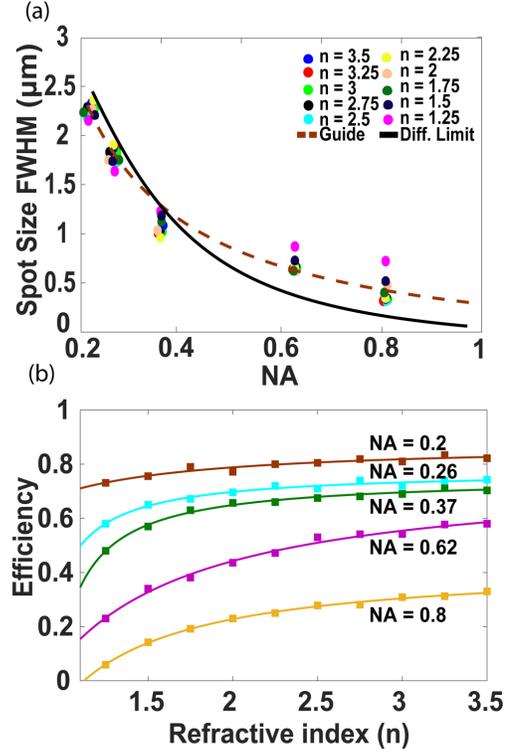

Figure 5. Performance of the metalens using the inverse design method: (a) The FWHM of the focal spot as a function of numerical aperture of the lens. The solid and dashed curves are the diffraction-limited FWHM, and guide to the eye, respectively. (b) shows the focusing efficiencies of metalenses with different numerical apertures against material refractive indices in the 1.25 to 3.5 range using the inverse design method. Different numerical apertures are specified with different colors and plotted curves are guides to the eye.

## 3. Discussion

A reasonably consistent behavior is observed between forward design (Figure 3) and inverse design (Figure 5) regarding the relation between focusing efficiency and the FWHM of the focal spots of the metalenses with the refractive index. For lower numerical apertures, there is no significant difference in focusing efficiency of metalenses with refractive indices in range of 1.25 to 3.5. However, for higher numerical apertures, getting better focusing efficiency of metalenses is feasible only by increasing their refractive index. Also, the FWHM of focal spots in the inverse design method are smaller than those in the forward design method. We attribute this to the choice of our figure of merit, in which we did not enforce a condition on the beam spot size.

The performance of a metalens is related to a metasurface beam deflector, whose performance as a function of material refractive index was recently analyzed at optical frequencies [16]. It was shown that, while over 80% efficiency can be obtained using high contrast materials such as Si and Ge, for large-angle ($> 60^o$) beam deflection, the efficiency drops quickly for low index materials, such as SiN. However, no significant difference in efficiency for modest deflection angles ($< 40^o$) was observed for different materials. this finding is consistent with our result that for high numerical aperture metalenses, higher refractive index materials provide higher efficiency.

## 4. Conclusion

We have evaluated low-loss dielectric materials with a wide range of refractive indices for designing metalenses using both forward and inverse design methodologies. We found reasonable agreement between both methods in terms of the focusing efficiency and the FWHM of the focal spots on the material refractive indices. We found that for low numerical apertures ($NA < 0.6$), the efficiency of the metalenses is almost independent of the refractive index. For higher numerical apertures, however, high-index materials provide higher efficiency. The relationship between refractive index and metalens performance is significant in choosing appropriate material, based on considerations like ease and scalability of manufacturing, or better tunability. In addition, we show that, even with very low refractive index ($n < 2$), we can achieve reasonable efficiency in a metalens, which will be important for printable photonics technologies.

**Funding.** The research is supported by the Samsung Global Research Outreach (GRO) grant, and the UW Reality Lab, Facebook, Google, and Huawei.

**Acknowledgment.** We thank NVIDIA corporation for the contribution of the GPU for our calculations in the inverse design method. In addition, we thank Amos Egel, Lorenzo Pattelli, and Giacomo Mazzamuto for allowing us to use and further contribute to CELES.